# Nanomagnetic Logic and Magnetization Switching Dynamics in Spin Torque Majority Gates

Dmitri E. Nikonov[1], *Senior Member, IEEE*, George I. Bourianoff[2], *Member, IEEE*, Tahir Ghani[3], *Fellow, IEEE*, and Ian A. Young[1], *Fellow, IEEE*

[1]Components Research, Intel Corp., Santa Clara, California 95052, USA
[2]Components Research, Intel Corp., Austin, Texas 78746, USA
[3] Portland Technology Development, Intel Corp., Hillsboro, Oregon 97124, USA

**Spin torque majority gates are modeled and several regimes of magnetization switching (some leading to failure) are discovered. The switching speed and noise margins are determined for STMGs and an adder based on it. With switching time of 3ns at current of 80μA, the adder computational throughput is comparable to that of a CMOS adder.**

## I. Spintronic Logic

SPINTRONIC devices find their main application in non-volatile memories, namely magnetic random access memory (MRAM). Recently, magnetic memory based on switching by injection of spin-polarized current, spin transfer torque RAM (STTRAM) [1], proved to be many times more efficient than previous types of MRAM. It is natural to extend the physics of spintronics to logic devices [2]. The expected benefits are reconfigurable and non-volatile logic, which does not suffer from standby power dissipation and can be turned on instantly. In spite of numerous spintronic logic devices proposed, few of them have been fabricated and none were demonstrated to function in an integrated circuit. A spin logic device, a spin torque majority gate (STMG), has been proposed [3,4], which leverages well-developed processes and materials used in STTRAM. In the present work we reveal the character of magnetization dynamics in switching of this device, describe the operation of a practically important circuit, a one bit of a full adder, and obtain performance projections for it. These results support feasibility of experimental implementation of STMG, demonstrate the possibility of creating extended spintronic circuits (e.g. adders) without the need of spin-to-electrical conversion, and provide an argument that such circuits can have performance comparable with the incumbent CMOS technology.

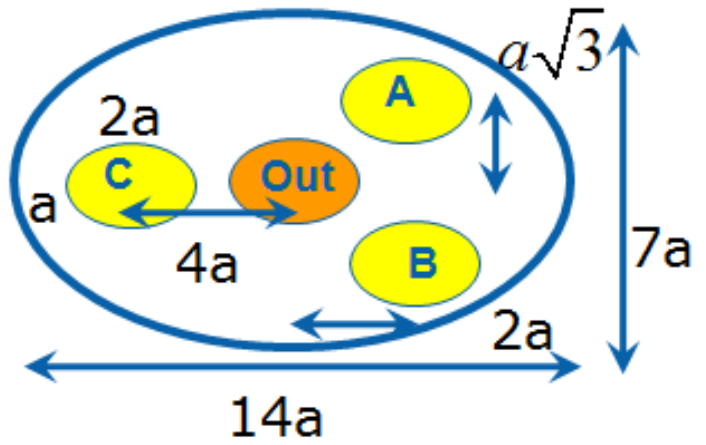

Fig 1. Layout of STMG. Input nanopillars are "A", "B" and "C", output pillar is "Out" in the middle. Minimum width is *a*. The aspect ratio for all ellipses is 2.

## II. Magnetization Dynamics

The structure of the STMG device with in-plane magnetization is shown in Figs. 1 and 2. The stack of the layers is similar to a magnetic tunnel junction (MTJ).

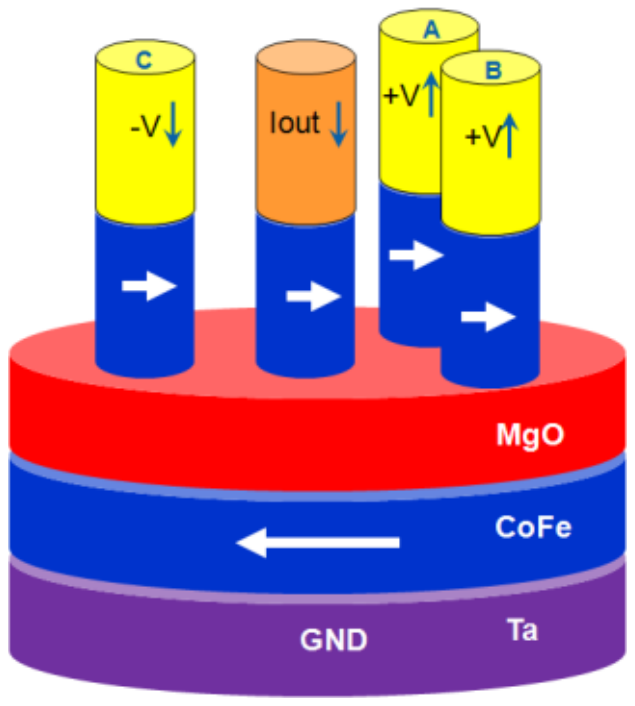

Fig 2. Scheme of STMG layers. Every nanopillar has its own fixed FM layer and a metal contact on the top. The common free FM layer, thickness *t*, below, is separated by a tunneling barrier of MgO.

The combined action of spin torques [5] resulting from currents in the three input pillars transfers enough torque to switch magnetization in the common free ferromagnetic (FM) layer. Digital inputs, which are encoded as voltage polarities designated as plus (p) or minus (m), determine the directions of torques. The three torques fight to force the magnetization direction in the free layer, which is in the end set by the majority of them.

This magnetization of the free layer is sensed with a sense amp via the tunneling magnetoresistance (TMR) effect [6] measured at the center pillar. A single STMG has a useful functionality as a reconfigurable AND/OR gate.

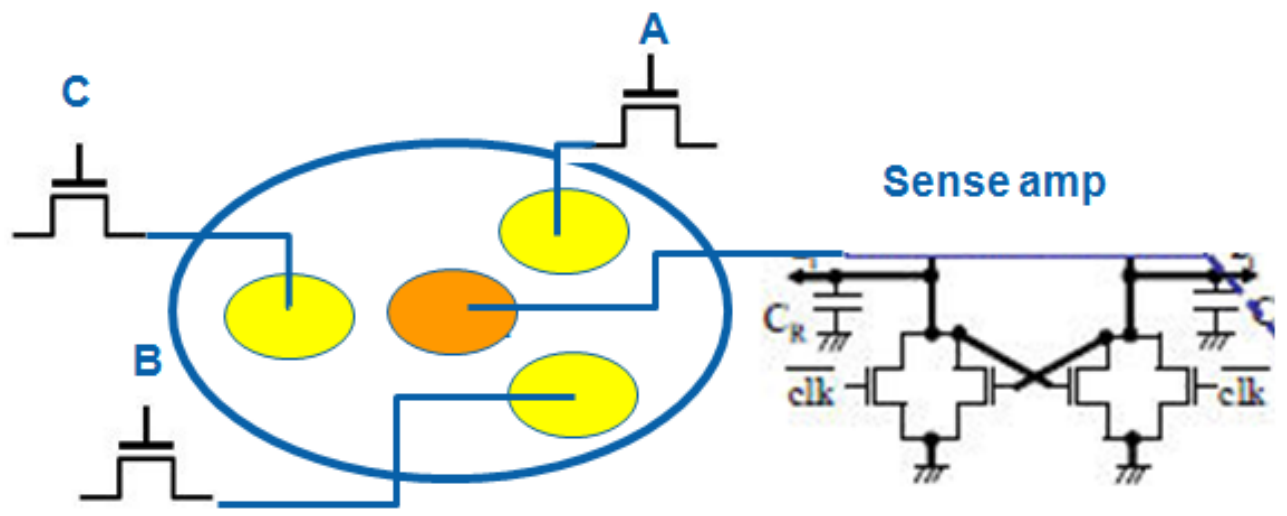

Fig. 3. Reconfigurable AND/OR gate with STMG. A transistor drives each input. Output detected by a sense amp.

The magnetization dynamics were modeled with OOMMF solver [7]. It is based on the solution of Landau-Lifshitz-Gilbert equations with the magnetization varying over the coordinate in plane of the device. In these simulations, the random thermal fluctuations of magnetization are neglected, which corresponds to zero effective temperature. We model cases when the initial magnetization is uniformly pointing to the right (average relative value 1, in units of saturation magnetization $M_s$), along the easy axis of the ellipse. The expected result of spin torque switching is for magnetization to end up pointing to the left (average relative value -1). Current in each pillar is indicated.

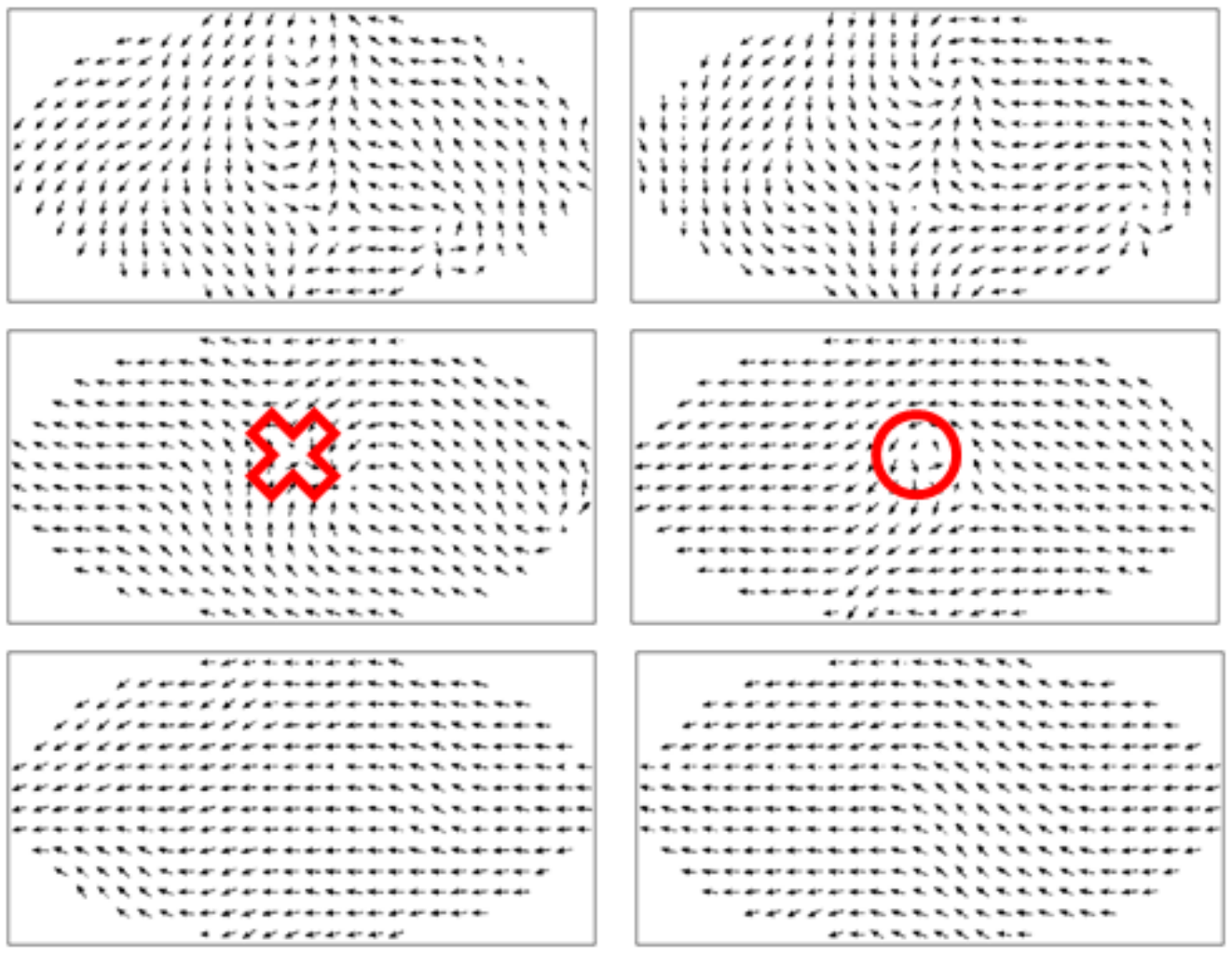

Fig 4. Magnetization patterns with polarities (ppp), $I$=4mA, $a$=24nm, $t$=2nm, at time intervals 0.2ns, left to right in rows. A vortex highlighted by a circle, an antivortex – by a cross.

The dynamics of magnetization proves to exhibit complex geometrical patterns. It may reach the desired final state without (Fig. 5) or with (Fig. 4) transient formation of vortices and anti-vortices. In other cases the final state may contain a vortex (Fig. 6) or and anti-vortex (Fig. 8), with a possibility of an anti-vortex forming and then disappearing at the edge of the free layer (Fig. 7). A vortex or an anti-vortex in the final magnetization state results in a failure of majority logic. In this case the magnetization under the output pillar is not along the easy axis and will not provide the right resistance value to sense. Such a situation must be avoided by a proper choice of geometry and current magnitude and duration. In further plots we mark such failure cases as zero switching speed. Note that they happen over a limited range of parameters. It is possible to find a broad operation range where normal switching occurs for all polarities of inputs.

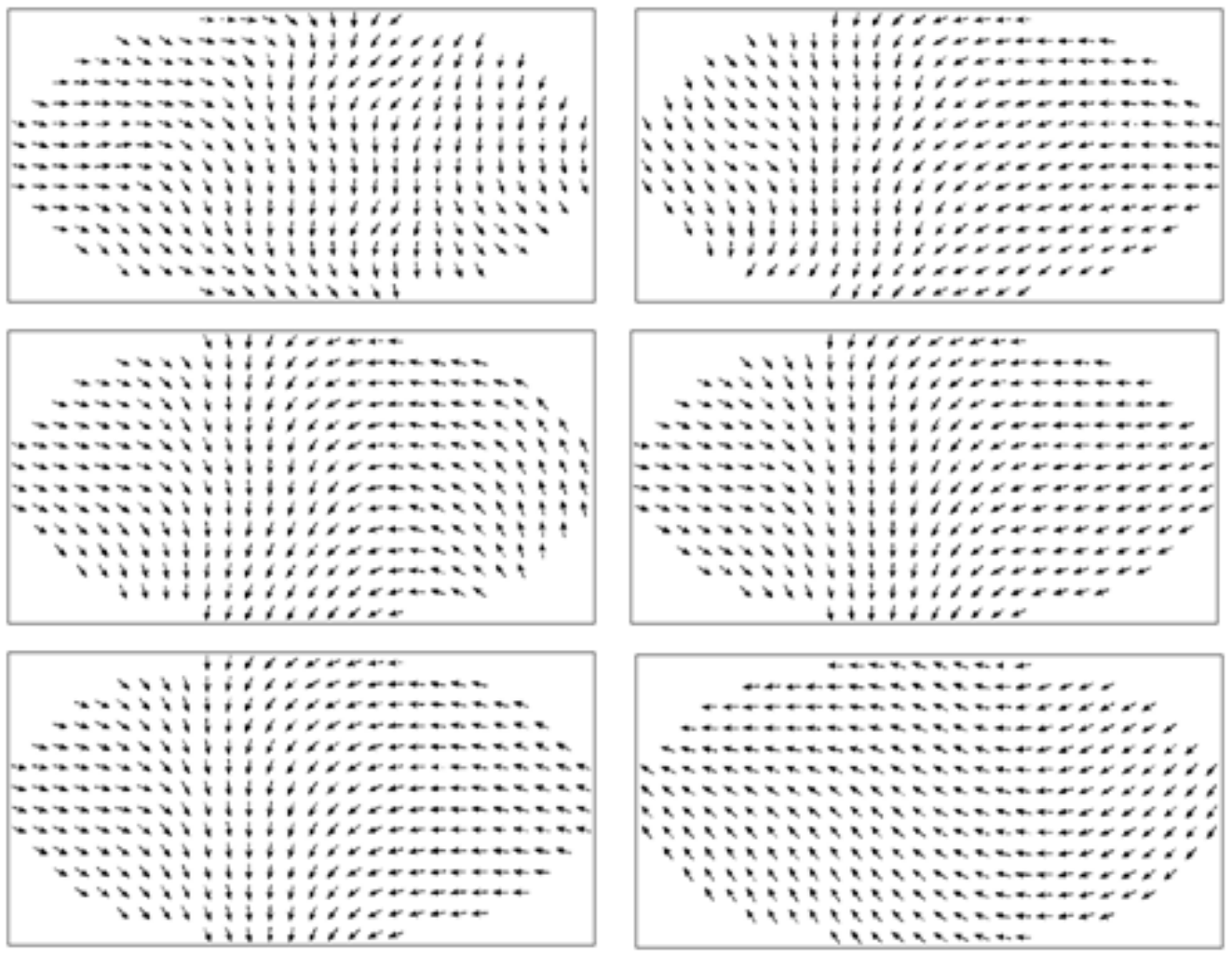

Fig 5. Same as Fig. 4, but polarities (ppm).

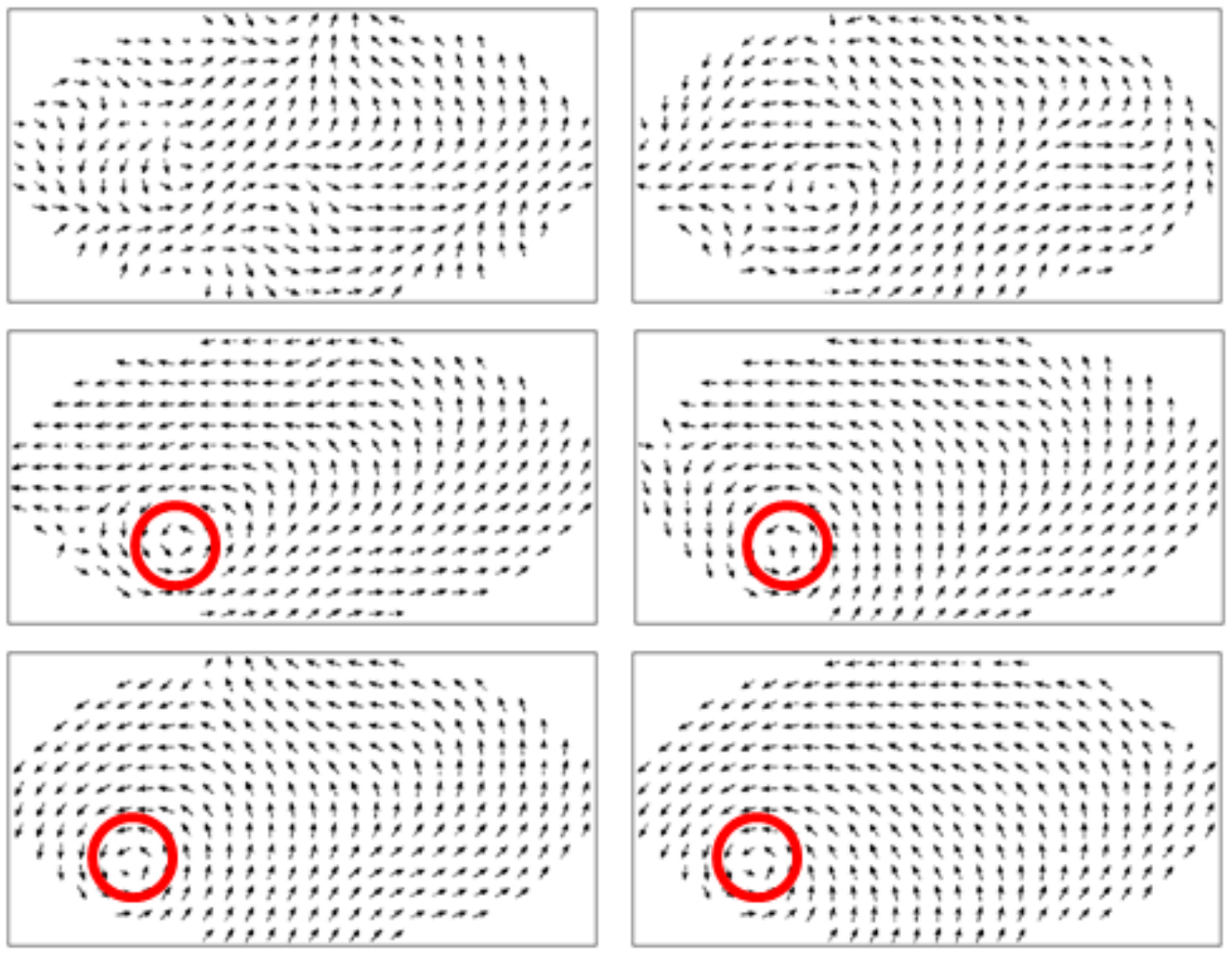

Fig 6. Same as Fig. 4, but polarities (pmp).

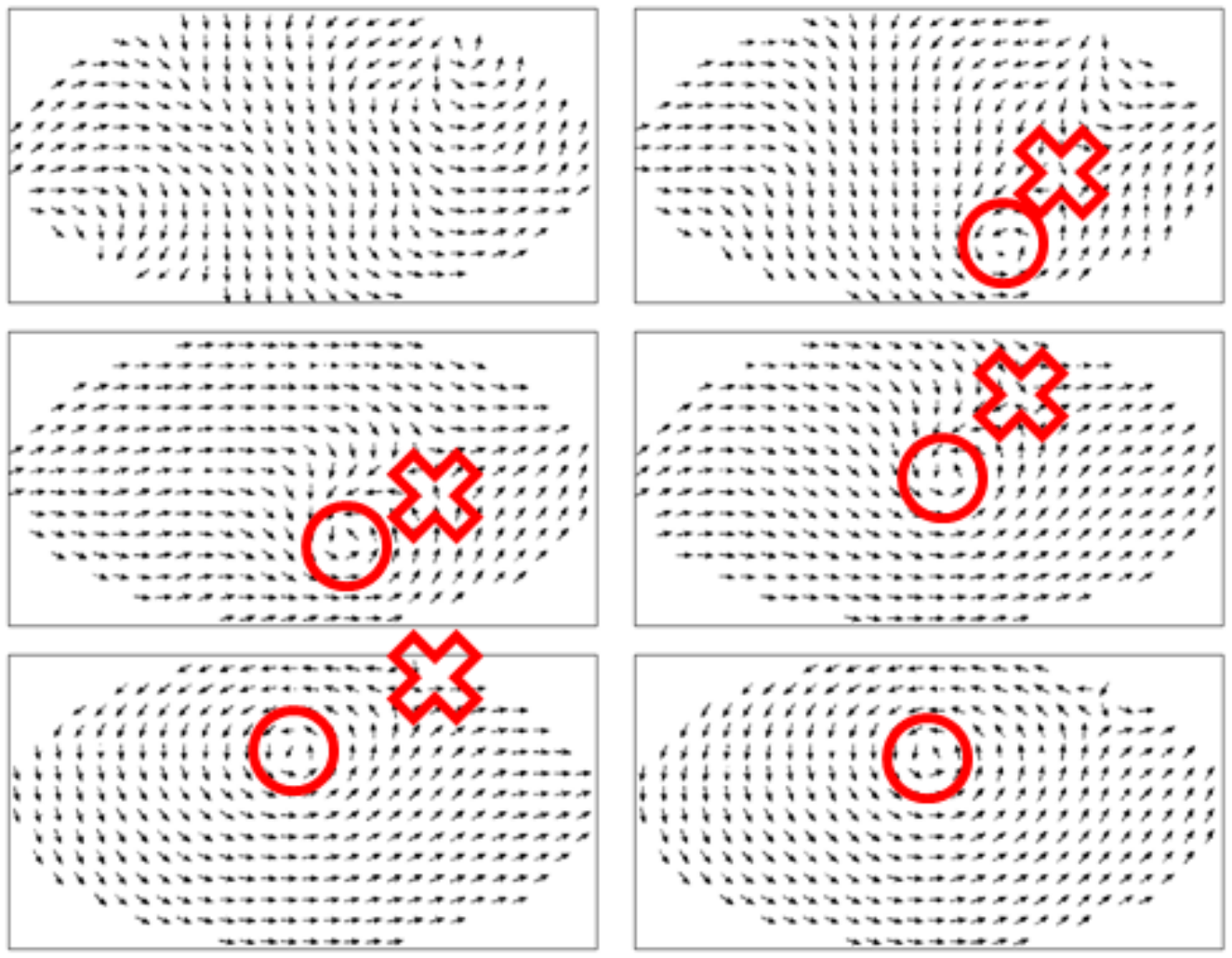

Fig 7. Magnetization patterns with polarities (ppm), $I$=16mA, $a$=48nm, $t$=3nm, at 0.2ns and then time intervals of 0.6ns.

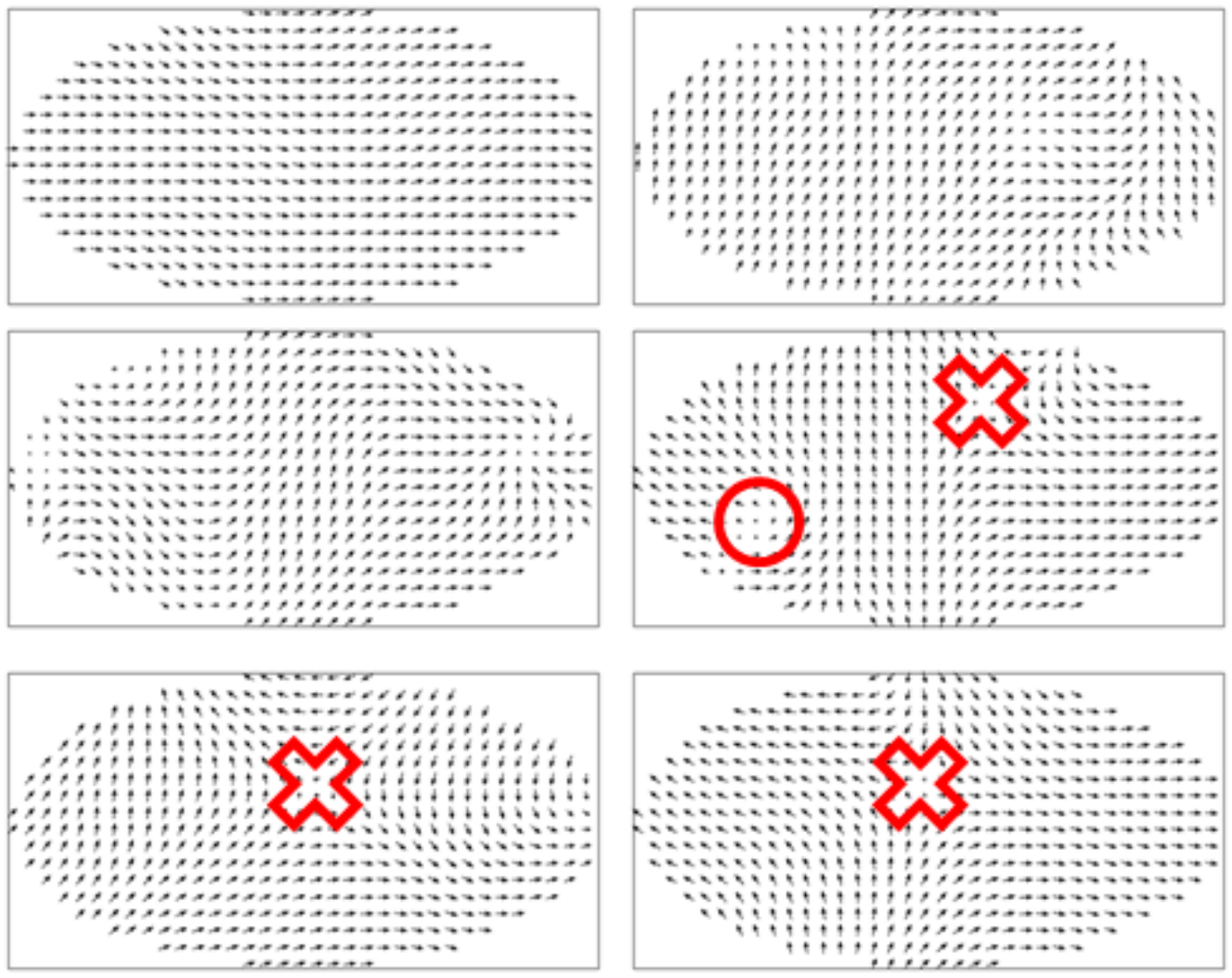

Fig 8. Magnetization patterns with polarities (pmp), $I$=1mA, $a$=12nm, $t$=2nm, at time intervals of 0.2ns.

## III. SWITCHING PERFORMANCE

Simulations at various current values and size provide values of switching speed (Fig. 9), which we define as the inverse time from the onset of the current pulse to the last instant when the average relative magnetization crosses −0.8. At lower values of current, the spin torque is not enough to overcome damping. So STMG has a threshold current similarly to STTRAM. Switching also fails (designated as zero speed) at larger values of $a$=24nm and for larger values of current due to a vortex formation. Even in this case, normal switching occurs for a broad range of intermediate value of current.

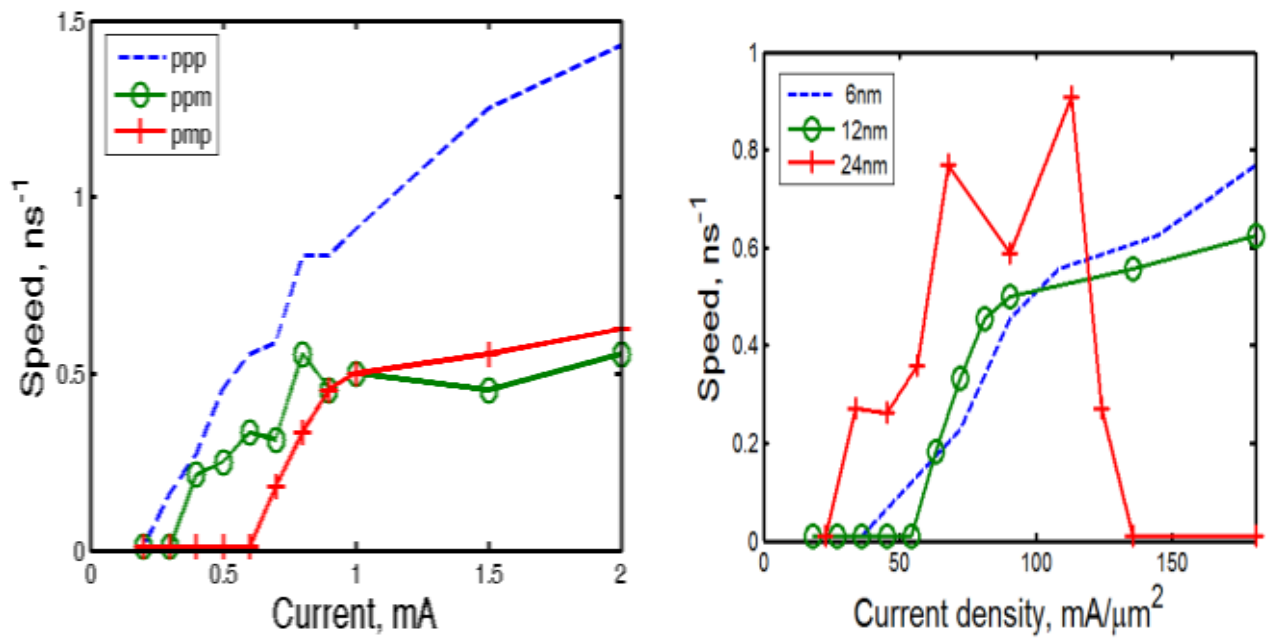

Fig 9. Switching speed of STMG (left) vs. current for various voltage polarities, $a$=12nm, $t$=2nm; (right) vs. current density per unit area of free layer for (pmp) polarity at various $a$.

An important requirement to logic is that noise in the input should be suppressed and not affect the output. Thermal noise in magnetic circuits appears as fluctuations of magnetization direction. We perform simulations of STMG switching with directions of magnetizations in two input fixed layers kept constant. The direction of magnetization in the fixed layer of the third input is varied from 0deg (pointing to the right) to 180deg (pointing to the left). We obtain the magnetization in the free layer of the output (Fig. 10). This shows that the output angle obeys a sharply non-linear transfer characteristic, similar to that of CMOS inverter, with a noise margin of ~84$^o$. The slope in the mid-point corresponds to gain of ~15.

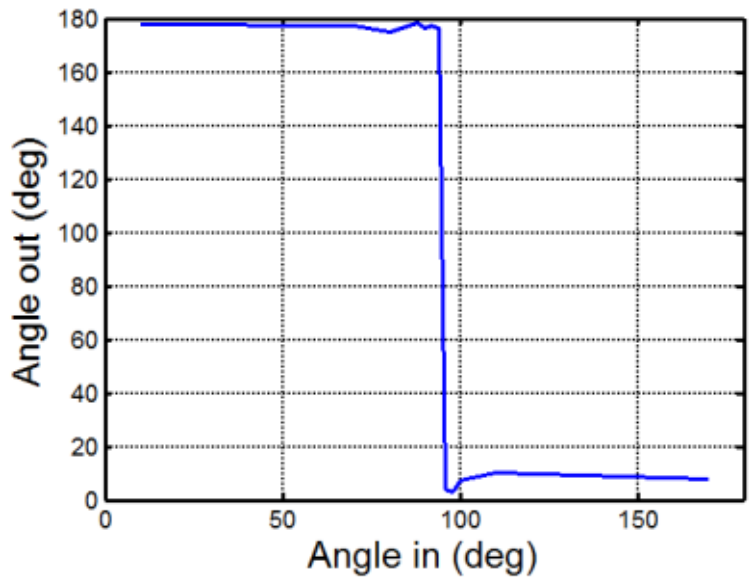

Fig 10. Angle of final magnetization vs. angle of spin polarization in nanopillar A, for polarization (pmp), $I$=1mA, $a$=12nm, $t$=2nm.

STMGs can also be fabricated from materials with out-of-plane magnetization (such as FePt or TbCoFe) and have the shape of crosses rather than ellipses. One expects lower switching current with perpendicular magnetization, see e.g. [8] but the structure of the layers (Fig 11) is more complicated. Both the free and the fixed layers are formed as synthetic antiferromagnets (SAF), consisting of two ferromagnetic layers (such as CoFe) separated by a thin metal layer (such as 0.8nm off Ru). This is done to permit a simple passive element to perform and inverter function: if the top of the SAF on one side is connected to the bottom of SAF on the other side, the magnetizations in the top layer, sensed in TMR, are opposite.

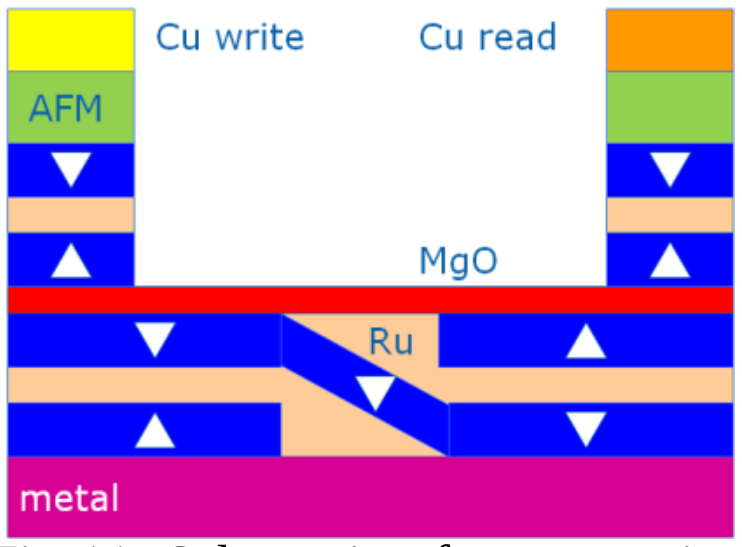

Fig 11. Schematic of cross-section of layers for devices with out-of-plane magnetization. The middle section is an inverter.

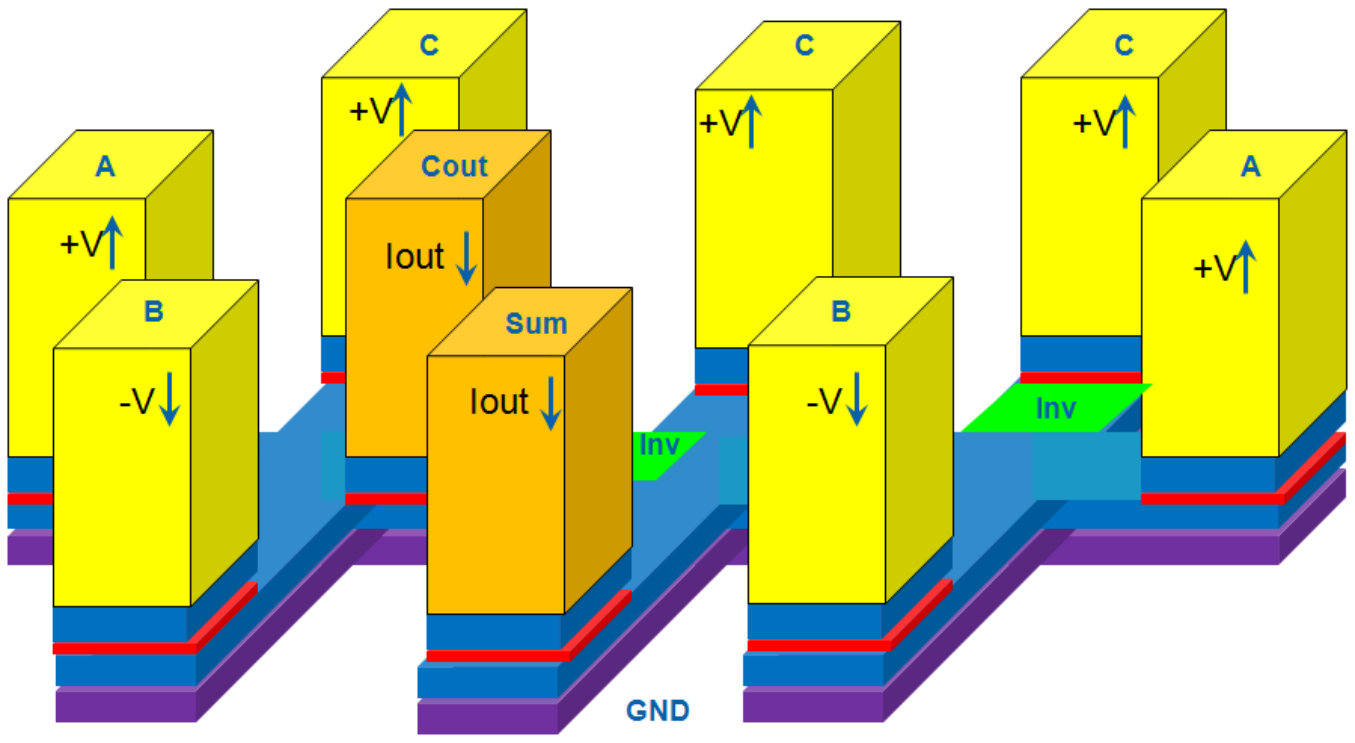

Fig 12. Schematic of an adder, width of FM wires is $a$.

Cross-shaped STMGs may also be concatenated to form more complex magnetic circuits. E.g. a magnetic adder [3] can be formed with three such STMG crosses of FM wires (Fig. 12). Two inverters are necessary to implement it.

Our simulations prove that the adder has the correct functionality. Using out-of-plane magnetization indeed allows a smaller value of the switching current (Fig. 13) (similarly to STTRAM.) For the wire width of 20nm, the switching time ~3ns can be achieved with a relatively small current of 80μA.

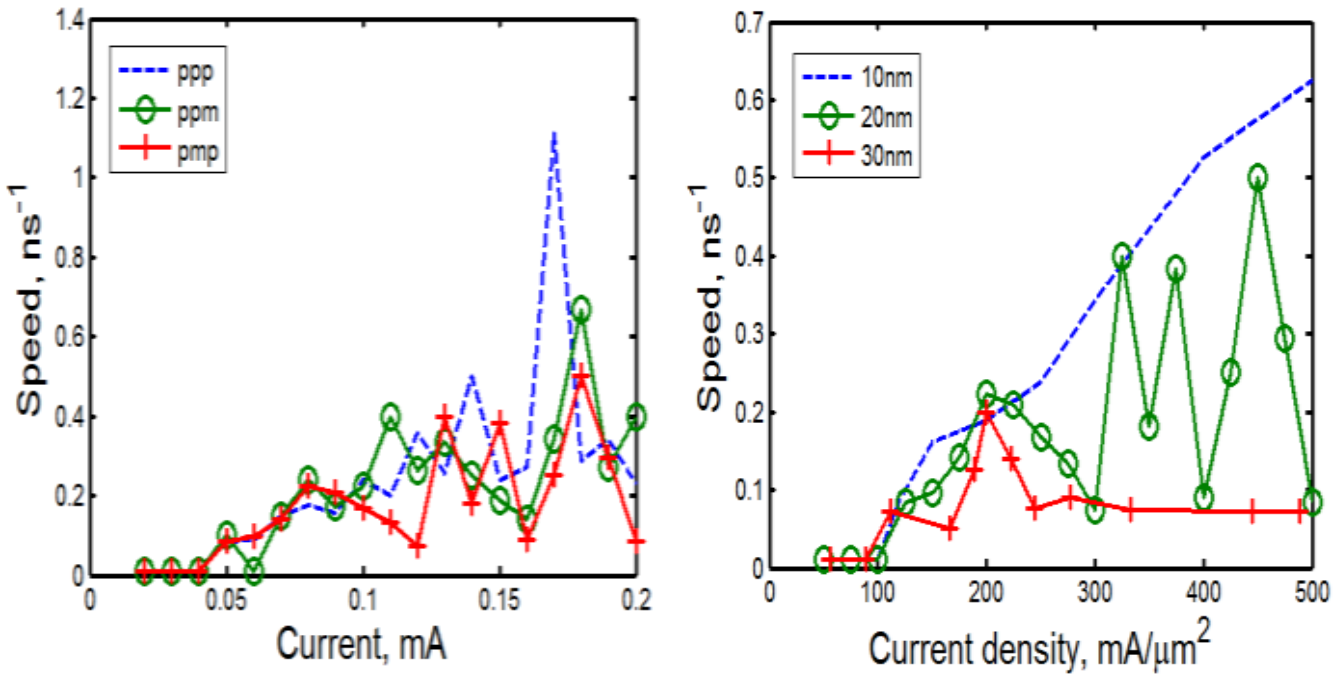

Fig 13. Switching speed in the adder (left) vs. current for various voltage polarities at *a*=20nm, *t*=2nm; (right) vs. current density per unit of nanopillar area for (pmp) polarity at various *a*.

By choosing an operation point from these plots, we can estimate the circuit's performance. In Fig. 14 we compare the STMG adder, the adder based on standard CMOS [9], and an adder composed of MTJs and CMOS. This last adder is based on a design and parameters in [10] scaled from the generation of 180nm to 22nm. The scaling of the delay was assumed proportional to the size, and the switching energy proportional to the square of the size. All three circuits were adjusted to work at approximately the same power per unit area.

| | CMOS | STMG | MTJ+CMOS |
|---|---|---|---|
| Process feature, F, nm | 22 | 22 | 22 |
| Area factor, F*F | 10413 | 2727 | 10124 |
| Area per gate, $um^2$ | 5.0 | 1.3 | 4.9 |
| Voltage, V | 0.81 | 0.1 | 0.81 |
| Switching time, ps | 16 | 2826 | 1.25 |
| Clocking time, ps | 250 | 5651 | 2000 |
| Switching energy intrinsic, aJ | 1382 | 147703 | ??? |
| Switching energy with circuits, aJ | 17640 | 257680 | 326000 |
| Power per gate, active, uW | 70.6 | 45.6 | 163.0 |
| Power per gate, standby, uW | 0.81 | 0 | 0 |
| Activity factor | 0.01 | 0.01 | 0.01 |
| Power per gate, average, uW | 1.52 | 0.46 | 1.63 |
| Power per unit area, $W/cm^2$ | 30.1 | 34.5 | 33.3 |
| Throughput, $Mops/ns/cm^2$ | 79.4 | 13.4 | 10.2 |

Fig 14. Table of comparison of performance of adder created with different technologies.

## IV. Conclusion

We conclude that STMG has ~6x smaller throughput than CMOS, but this disadvantage is offset by advantages of non-volatility and reconfigurability.

## Acknowledgment

The authors thank Robert McMichael and Michael Donahue for providing a version of NIST's OOMMF software [7].